\documentclass[aps,twocolumn,showpacs,preprintnumbers,nofootinbib,prd,superscriptaddress,groupedaddress,10pt]{revtex4-1}

\makeatletter
\def\l@subsubsection#1#2{}
\def\l@subsubsubsection#1#2{}
\makeatother

\usepackage{graphicx,amssymb,amsmath,amsthm,amsfonts,epsfig,epsf,fixmath}
\usepackage[usenames]{color}
\usepackage{epstopdf}

\usepackage{bm}
\usepackage{amsmath}
\usepackage{dcolumn}
 \usepackage[latin1]{inputenc}
\usepackage{latexsym}
\usepackage{rotating}
\usepackage{longtable}

\usepackage{enumerate}
\usepackage{tensor,multirow}
\usepackage{mathtools}
\usepackage{url}
\usepackage[linktocpage]{hyperref}

\def\nn{\nonumber}

\begin{document}

\title{Tidal heating as a discriminator for horizons in extreme mass ratio inspirals}

\author{
Sayak Datta$^1$,
Richard Brito$^2$,
Sukanta Bose$^{1,3}$,
Paolo Pani$^{2,4}$,
Scott A.\ Hughes$^5$
}

\affiliation{$^{1}$ Inter-University Centre for Astronomy and Astrophysics, Post Bag 4, Ganeshkhind, Pune 411 007, 
India}

\affiliation{$^{2}$ Dipartimento di Fisica, ``Sapienza'' Universit\`a di Roma, Piazzale 
Aldo Moro 5, 00185, Roma, Italy}

\affiliation{$^{3}$ Departmentof Physics \& Astronomy, Washington State University, 1245 Webster, Pullman, WA 
99164-2814, U.S.A.}

\affiliation{$^{4}$ Scuola Superiore di Studi Avanzati Sapienza, Viale Regina Elena 291, 00161, Roma, Italy}

\affiliation{$^{5}$ Department of Physics and Kavli Institute for Astrophysics and Space Research, Massachusetts 
Institute of Technology, Cambridge, Massachusetts 02139, USA}

\begin{abstract} 
The defining feature of a classical black hole is being a perfect absorber. Any evidence showing otherwise would 
indicate a departure from the standard black-hole picture. Energy and angular momentum absorption by the horizon of a 
black hole is responsible for tidal heating in a binary. This effect is particularly important in the latest 
stages of an extreme mass ratio inspiral around a spinning supermassive object, one of the main targets of the future 
LISA mission. We study how this effect can be used to probe the nature of supermassive objects in a model independent 
way. We compute the orbital dephasing and the gravitational-wave signal emitted by a point particle in 
circular, equatorial motion around a spinning supermassive object to the leading order in the mass ratio. Absence 
of absorption 
by the central object can affect the gravitational-wave signal dramatically, especially at high spin. 
This effect will make it possible to put an unparalleled upper bound on the reflectivity of exotic compact 
objects, at the level of ${\cal O}(0.01)\%$. This stringent bound would exclude the possibility of
observing echoes in the ringdown of a supermassive binary merger.
\end{abstract}

\maketitle

\section{Introduction}
At the classical level, black holes~(BHs) in general relativity are perfect absorbers since their defining 
characteristic --~the event horizon~-- is a one-way, null hypersurface.
Measuring some amount of reflectivity near a dark compact object would be a smoking gun 
of departures from the classical BH picture~\cite{Cardoso:2019rvt}.
Although modelling the reflectivity of exotic compact objects~(ECOs) is challenging (see Ref.~\cite{Oshita:2019sat} for 
recent progress in a specific model), the absence of a horizon or the presence of some nearby structure would 
necessarily imply imperfect absorption. 
Thus, searching for this effect provides a model-independent test of ECOs and could help quantify the ``BH-ness'' of a 
dark compact object, e.g. by placing an upper bound on its reflectivity.

A spinning BH absorbs radiation of frequency $\omega>m\Omega_H$ (where $m$ is the azimuthal number of the 
wave and $\Omega_H$ is the BH angular velocity) but amplifies radiation of smaller frequency, due to 
superradiance (see~\cite{Brito:2015oca} for a review).
The combination of these absorbing and amplifying behaviors means that BHs are dissipative systems which behave like a 
Newtonian viscous fluid~\cite{MembraneParadigm,Damour_viscous,Poisson:2009di,Cardoso:2012zn}. 
Dissipation gives rise to various interesting effects in a binary system --~such as tidal 
heating, tidal acceleration, and tidal locking, as in the Earth-Moon 
system, where dissipation is provided by the friction of the oceans with the crust.

The members of a binary feel each others' tidal fields particularly strongly late in the inspiral, as the bodies 
approach their final plunge and merger.  If the bodies are (at least partially) absorbing, these tides backreact on the 
orbit, transferring energy and angular momentum from their spin into the orbit.  This effect is called {\it tidal 
heating}~\cite{Hartle:1973zz,PhysRevD.64.064004,PoissonWill}. 
Tidal fields on BHs satisfy a unique boundary condition which picks out how a BH's spin is transferred to 
the orbit. Tidal heating can be responsible for thousands of radians of accumulated 
orbital phase~\cite{Hughes:2001jr,Bernuzzi:2012ku,Taracchini:2013wfa,Harms:2014dqa,Datta:2019euh} for 
extreme-mass-ratio inspirals~(EMRIs) in the band of the future 
space-based Laser~Interferometer~Space~Antenna~(LISA)~\cite{Audley:2017drz} and of evolved concepts 
thereof~\cite{Baibhav:2019rsa}. This large effect is due to the dissipative nature of BH horizons, and allows 
for rather exquisite tests of the nature of supermassive objects.

If at least one binary member is an ECO instead of a BH, the dissipation is likely to be much smaller, 
even negligible, potentially changing the inspiral phase by a large amount, especially if the binary's members spin 
rapidly. Therefore, even in those cases in which the external geometry of the ECO is extremely close to that of a Kerr 
BH, tidal heating can provide a powerful and model-independent discriminator for the existence of horizons and for the 
nature of supermassive objects~\cite{Maselli:2017cmm, Datta:2019euh}. This adds to other EMRI-based tests, namely no-hair theorem tests 
based on measurements of the quadrupole moment of the central object~\cite{Ryan:1995wh,Barack:2006pq,Babak:2017tow, Datta:2019euh}, 
and null-hypothesis tests based on the absence of tidal Love numbers~\cite{Pani:2019cyc}. Altogether, these tests 
suggest that EMRIs will be unique probes of the nature of supermassive objects (for recent reviews on these and 
other tests, see Refs.~\cite{Cardoso:2019rvt,Baibhav:2019rsa}).

A detailed calculation is needed to determine how tidal heating would work for an ECO~\cite{Maselli:2017cmm}, and the 
answer will necessarily depend on the specific ECO model~\cite{Oshita:2019sat}. However, 
by losing the horizon boundary condition, it is certain that the tidal coupling of the orbit to the object will change. 
A high signal-to-noise ratio~(SNR) measurement should be able to determine the impact of this effect with 
unparalleled precision, either for EMRIs around highly-spinning supermassive 
objects~\cite{Hughes:2001jr,Datta:2019euh}, or for  highly-spinning, supermassive binaries~\cite{Maselli:2017cmm}. 

The goal of this paper is to quantify this expectation. In particular, we wish to estimate the projected constraints 
on the reflectivity of a spinning supermassive 
object that would arise from measuring the tidal heating in an EMRI.

Overall, even making the conservative assumption that the geometry around the object can be approximated with 
that of a Kerr BH (as suggested by various arguments~\cite{Raposo:2018xkf,Barcelo:2019aif}, see next section), the 
absence of a horizon would produce three main effects in the inspiral:
\begin{itemize}
 \item \emph{Boundary conditions} for radiation near the surface of the object would be different.
 \item As a result of the above, the quasinormal modes of the object would differ from those of Kerr. In 
particular, low-frequency modes generically emerge~\cite{Pani:2009ss,Maggio:2017ivp,Maggio:2018ivz}, which might be
\emph{resonantly excited} during the 
inspiral~\cite{Pani:2010em,Macedo:2013jja,Cardoso:2019nis}. 
 \item Again as a result of different boundary conditions near the surface, at least part of the radiation is 
reflected back, providing at least some \emph{reflectivity}.
\end{itemize}
Clearly, the boundary conditions are model dependent, and so are the quasinormal-mode frequencies. Furthermore, the 
effect of resonances has been recently investigated and was shown to be negligible, at least for non-spinning 
ECOs~\cite{Cardoso:2019nis}.
On the other hand, partial reflectivity is a necessary and generic prediction of the absence of a horizon and can be 
constrained in a model-independent way. Understanding the consequences of this fact will be our focus in this analysis.

\section{Setup}
Henceforth we use $G=c=1$ units. We shall denote the mass and angular momentum of the central object by $M$ and 
$J=aM=\chi M^2$, respectively. The mass of the small orbiting (nonspinning) body is $\mu$ and the mass ratio is denoted 
by $\nu=\mu/M\ll1$.

\subsection{Background}
We consider a spinning compact object whose exterior geometry is described by the Kerr 
metric~\cite{Maggio:2017ivp,Abedi:2016hgu,
Wang:2018gin, Barausse:2018vdb,Maggio:2018ivz}. 
Unlike the case of spherically symmetric spacetimes, the absence of Birkhoff's theorem in axisymmetry 
does not ensure that the vacuum region outside a spinning object is described by the Kerr geometry. This implies that 
the multipolar structure of a spinning ECO might be different from that of a Kerr 
BH~\cite{Raposo:2018xkf,Barcelo:2019aif}. Nevertheless, for perturbative solutions to the vacuum Einstein's equation 
that admit a smooth BH limit, all multipole moments of the external spacetime approach those
of a Kerr BH in the high-compactness regime~\cite{Raposo:2018xkf} (for specific examples, 
see Refs.~\cite{Pani:2015tga,Uchikata:2015yma,
Uchikata:2016qku,Yagi:2015hda,Yagi:2015upa,Posada-Aguirre:2016qpz}).
Therefore, we conservatively assume that the small object follows the geodesics of a Kerr metric, with orbital 
parameters that evolve secularly due to energy and angular momentum fluxes. These fluxes might be different if the central 
object is a BH or an ECO, as discussed below.

In Boyer-Lindquist coordinates, the line element outside the object reads
\begin{eqnarray}
ds^2&&=-\left(1-\frac{2Mr}{\Sigma}\right)dt^2+\frac{\Sigma}{\Delta}dr^2-\frac{
4Mr}{\Sigma}a\sin^2\theta d\phi dt  \nn \\
&+&{\Sigma}d\theta^2+
\left[(r^2+a^2)\sin^2\theta +\frac{2Mr}{\Sigma}a^2\sin^4\theta
\right]d\phi^2\,.\label{Kerr}
\end{eqnarray}
In the above equation $\Sigma=r^2+a^2\cos^2\theta$ and $\Delta=r^2+a^2-2M r = (r - r_{+})(r - r_{-})$, where $r_{\pm}=M 
\pm \sqrt{M^2-a^2}$. The angular velocity at the event horizon is $\Omega_H = \chi/(2r_+)$.

We shall assume that the object is as compact as\footnote{Our results are based only on the fact that the 
geometry outside the innermost-stable circular orbit~(ISCO) is described sufficiently well by the Kerr metric. Indeed, 
after the small body crosses the ISCO it plunges directly, and the signal emitted during the plunge is negligible 
compared to the rest of the inspiral.} a Kerr BH, i.e. its radius is close to $r_+$.  The properties of the object's 
interior and 
surface can be parametrized in terms of the fraction of radiation that is absorbed compared to the BH case, as 
discussed below.

\subsection{Linear perturbations by a point-like source: the BH case}
%
In order to elucidate the differences relative to the case in which the central object is a Kerr-like ECO, we start by 
reviewing the case of a point-like source in circular, equatorial orbit around a Kerr BH.

The emitted gravitational radiation can be studied by solving the Teukolsky equation for spin $s=-2$ perturbations, 
which describes the curvature invariant $\psi_4$. The latter can be decomposed as
\begin{equation}
\label{multipola decomposition}
    \psi_4 = \frac{1}{(r-iM\chi\cos\theta)^4}\int_{-\infty}^{\infty}d\omega\sum_{\ell m}R_{\ell m \omega}(r)S_{\ell m 
\omega}(\theta,\phi) e^{-i\omega t}\,,
\end{equation}
where the sum runs over $\ell\geq2$ and $-\ell \leq m \leq \ell$. 
The function $S_{\ell m \omega}(\theta,\phi)$ is a spheroidal harmonic of spin weight $-2$. The radial function 
$R_{\ell m \omega}(r)$ satisfies the following equation,

\begin{equation}
\label{radial equation}
    \Delta^2 \frac{d}{dr}\left(\frac{1}{\Delta}\frac{d R_{\ell m \omega}}{dr}\right) - V(r)R_{\ell m \omega} = - 
\mathcal{T}_{\ell m \omega}(r)\,,
\end{equation}
where the potential $V(r)$ can be found, e.g., in Refs.~\cite{Drasco:2005kz, Hughes:1999bq, Cardoso:2019nis}. 
The source $\mathcal{T}_{\ell m \omega}(r)$ is constructed from certain projections of the energy-momentum tensor of a 
point-like source:
\begin{equation}
T_{\alpha\beta} = \frac{\mu u_{\alpha}u_{\beta}}{\Sigma \sin\theta (dt/d\tau)} \delta [r-r_o(t)]\delta[\theta-\theta_o(t)]\delta[\phi-\phi_o(t)].
\end{equation}
where the subscript ``o'' is used to label the coordinates of the orbiting 
body's worldline. In the current work we focus on circular equatorial orbits. Therefore, $\theta_o(t) = \pi/2$ 
and $r_o(t) = r_{{\rm orbit}} =$ constant. The orbital radius is related to the orbital angular velocity $\Omega$ by 
$\Omega=M^{1/2}/(r_o^{3/2}+a M^{1/2})$.

We solve Eq.~\eqref{radial equation} by first building a Green's function from solutions of the homogeneous 
equation, and then integrating that function over the source~\cite{Drasco:2005kz, Hughes:1999bq} (see also Appendix D of 
\cite{OSullivan:2014ywd}, which translates the notation in this past work to the form that has recently been adopted by 
the BH perturbation theory community). The resulting solution has the following asymptotic behavior
\begin{equation}
    R_{\ell m \omega}(r) = 
    \begin{cases}
    Z^\infty_{\ell m \omega} e^{i\omega x} & r \rightarrow \infty \\
    Z^H_{\ell m \omega} e^{-ik x} & r \rightarrow r_+,
    \end{cases}
\end{equation}
where $k=\omega-m\Omega_H$, $x$ is the tortoise coordinate defined by
\begin{equation}
 \frac{dx}{dr}=\frac{r^2+a^2}{\Delta}\,,\label{tortoise}
\end{equation}
and
%
\begin{eqnarray}
Z^\infty_{\ell m \omega} &=& D^\infty \int_{r_+}^{\infty} dr' \frac{R^{\rm in}_{\ell m \omega}(r')\mathcal{T}_{\ell m 
\omega}(r')}{\Delta(r')^2}\,, \label{Zinfty}\\
%
    Z^{H}_{\ell m \omega} &=& D^{H} \int^{\infty}_{r_+} dr' \frac{R^{\rm up}_{\ell m 
\omega}(r')\mathcal{T}_{\ell m \omega}(r')}{\Delta(r')^2},
\end{eqnarray}
where $R^{\rm up, in}_{\ell m \omega}(r)$ are the homogeneous solutions of Eq.~\eqref{radial equation} with regular 
boundary conditions at infinity and at the horizon, respectively. The quantity $D^{\infty,H}$ is a shorthand notation 
for a collection of constants that can be found in Refs.~\cite{Drasco:2005kz,OSullivan:2014ywd}. 
If the orbits are periodic, then the spectrum of the coefficients $Z^{\infty,H}_{\ell m \omega}$ is discrete,
 \begin{equation}
     Z^{\infty,H}_{\ell m \omega} = Z^{\infty,H}_{\ell m} \delta(\omega - m\Omega)\,.
 \end{equation}
In this case the energy fluxes at infinity and at the horizon read
\begin{eqnarray}
 \dot E_\infty &=& \sum_{\ell m}\frac{|Z^{\infty}_{\ell m }|^2}{4\pi m^2\Omega^2}  \label{EdotINF} \\
 \dot E_H &=& \sum_{\ell m}\frac{\alpha_{\ell m}|Z^H_{\ell m }|^2}{4\pi m^2 \Omega^2}\label{EdotH}\,,
\end{eqnarray}
where $\alpha_{\ell m}$ is provided in Ref.~\cite{Taracchini:2013wfa}.
%
For circular and equatorial orbits, angular momentum fluxes are related to the energy fluxes by 
$\dot{E}_{\infty,H} = \Omega \dot{L}_{\infty,H}$.

In general $\dot E_H\ll \dot E_\infty$, although its relative importance grows with the BH spin and with $\Omega$. For 
example, for 
$\chi=0.998$, $|\dot E_H/\dot E_\infty|\approx 0.108$ at the ISCO.

\subsection{Modelling fluxes for a reflective ECO}\label{sec:absECO}

Let us now discuss how the above fluxes should be modified in case the BH horizon is replaced by a (partially) 
reflective ECO. We summarize here the main result; the detailed computation is given in Appendix~\ref{app:flux}

\subsubsection{Flux near the object}

The energy flux at the horizon can be expressed in terms of the fraction of energy 
absorbed by the 
object relative to the energy absorbed by the event horizon of a BH with the same mass and spin. 
Specifically, the flux of radiation across the ECO surface reads
\begin{equation}
 \dot E_{\rm ECO} = (1-|{\cal R}|^2) \dot E_H\,,  \label{EdotECO}
\end{equation}
where ${\cal R}$ is the reflectivity coefficient at the surface~\cite{Mark:2017dnq,Maggio:2017ivp}. This quantity can 
in general be frequency- and spin-dependent but for simplicity we will consider it to be constant. For a BH, 
${\cal R}=0$ whereas $|{\cal R}|=1$ for a perfectly reflecting object. Our goal is to place an upper bound on ${\cal 
R}$.

Regardless of the reflectivity, an ultracompact object can efficiently trap radiation within its photon 
sphere~\cite{Cardoso:2014sna,Cardoso:2016rao,Cardoso:2016oxy}, which mimicks the effect of a horizon. 
For example, suppose that the effective surface of the object is located at $r=r_+(1+\epsilon)$, with $\epsilon\ll1$. 
In the $\epsilon\to0$ limit we expect to recover the BH result, no matter the value of ${\cal R}$.

As we now show, trapping at the photon sphere is never effective in an EMRI system.  If radiation is trapped within the 
photon sphere for enough time, it is effectively lost from the energy balance.  This loss contributes to the orbital 
evolution. Whether this effect
is important can be quantified as follows~\cite{Maselli:2017cmm}. When $\epsilon\ll1$, the travel time for radiation is dominated by
the delay time near the surface of the object.  This travel time is (half of) the echo time scale~\cite{Cardoso:2016rao,Cardoso:2016oxy,Abedi:2016hgu},
and is given by 
\begin{equation}
 T_{\rm arr}\sim M\left(1+\left(1-\chi^2\right)^{-1/2}\right)|\log\epsilon|\,. \label{Tarr}
\end{equation}
Effective absorption occurs if the above time scale is much longer than a typical radiation-reaction time 
scale\footnote{Note that this argument revises that presented in Ref.~\cite{Maselli:2017cmm}.}, which we estimate as
\begin{equation}
 T_{\rm RR}\sim \frac{E}{\dot E_\infty}\sim \frac{5}{64}\left(\frac{r_o}{M}\right)^4 \frac{M}{\nu}\,. \label{TRR}
\end{equation}
Note that this is a leading-order estimate: $E= \nu M^2/(2r_o)$ is the binary's binding energy in Newtonian
gravity, and we used the quadrupole formula, $\dot E_{\infty} = (32/5)\nu^2(M/r_o)^5$ to estimate the GW flux.
Requiring $T_{\rm arr}\gg T_{\rm RR}$ yields the condition
\begin{equation}
 |\log\epsilon|\gg \frac{5}{64\left(1+\left(1-\chi^2\right)^{-1/2}\right)}\left(\frac{r_o}{M}\right)^4 \frac{1}{\nu}\,,
\end{equation}
Owing to the $1/\nu$ and $\log\epsilon$ dependence, this formula will never be satisfied in the EMRI limit,
except for  unrealistically small values of $\epsilon$.
In other words, for EMRI systems the radiation-reaction time scale is always so long that light-sphere trapping cannot provide 
effective absorption.  The only way for an ECO to absorb radiation is by dissipating within the object, as parametrized 
by Eq.~\eqref{EdotECO}.

\subsubsection{Flux at infinity}

Another important point concerns the energy flux at infinity.  From Eqs.~\eqref{Zinfty} and \eqref{EdotINF} we 
notice that $\dot E_\infty$ depends on the homogeneous solutions of Teukolsky's equation that is regular \emph{at the 
horizon} (cf.\ the dependence on the ingoing solution $R^{\rm in}_{\ell m\omega}$). Clearly that solution is different for an ECO, owing to the different boundary conditions. Nevertheless, in 
Appendix~\ref{app:flux}, we show that the energy flux at infinity is, up to numerical accuracy, the same for a BH or 
for an ECO, regardless of the reflectivity of the latter\footnote{In principle, there could be large effects very close 
to extremely narrow resonances~\cite{Pani:2010em}.  However, it has been shown that the impact of these
resonances is negligible~\cite{Cardoso:2019nis}, suggesting that the analysis in Appendix~\ref{app:flux}
(which ignores the resonances) is reliable.}.

To summarize, in order to study the adiabatic evolution of the EMRI to leading order in the mass ratio it is sufficient 
to compute the energy flux at infinity as in the BH case, and to account for (total or partial) absorption within the 
object using Eq.~\eqref{EdotECO}.

\begin{figure}[th]
\centering
\includegraphics[width=0.45\textwidth]{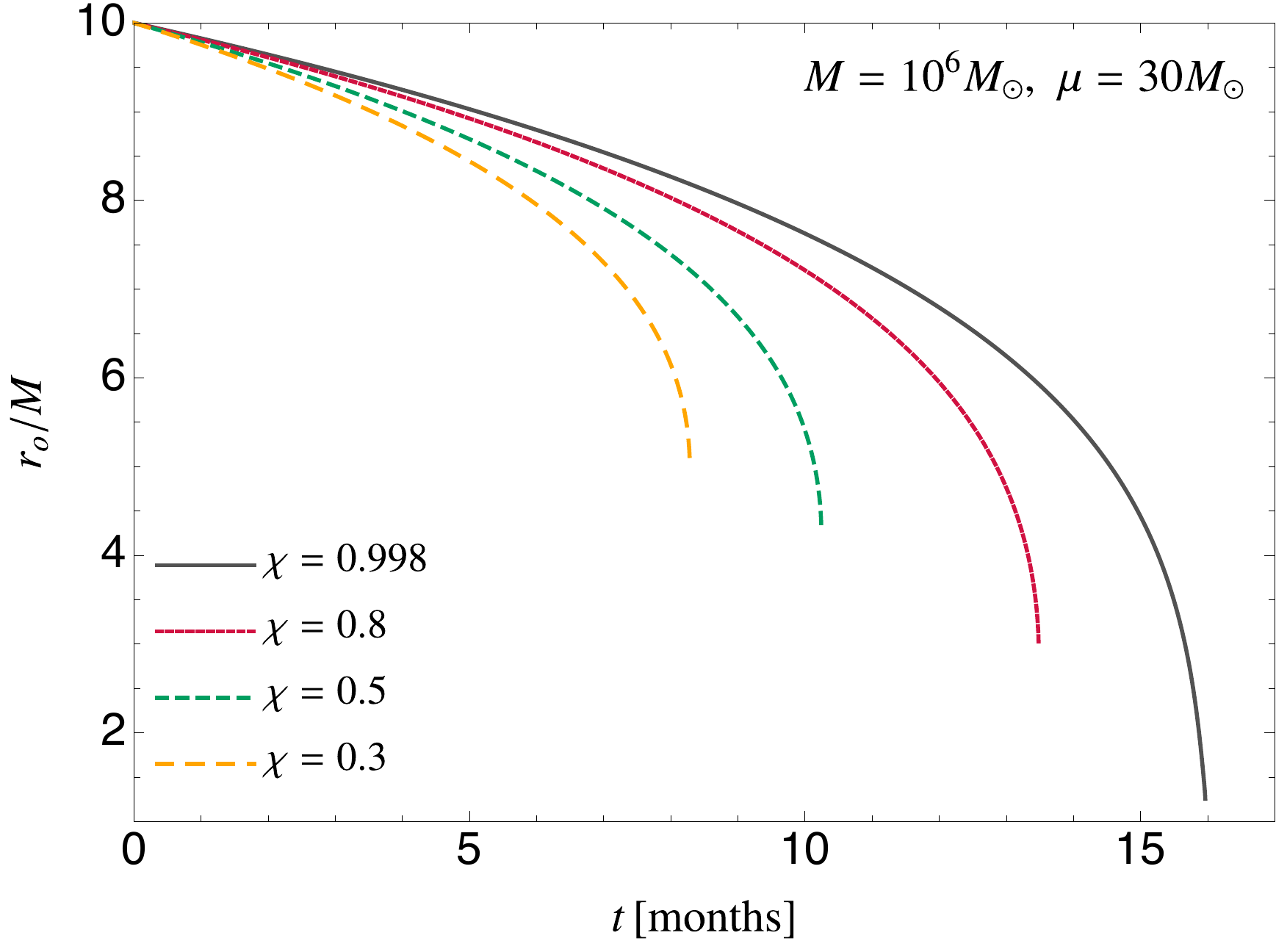}
\caption{Evolution of the orbital radius under radiation reaction when including the influence of tidal heating (i.e., when the central object is a perfect absorber, ${\cal R}=0$) for $M=10^6M_\odot$, $\mu=30M_\odot$ and various spin 
values. The evolution starts at $r_o=10M$ and ends at the ISCO.}
\label{fig:radius}
\end{figure}
%

\subsection{Circular equatorial orbits in Kerr and radiation reaction}

Circular equatorial orbits in Kerr can be uniquely parametrized in terms of the energy $E$ and the angular momentum $L_z$ of the small orbiting body given by:
\begin{eqnarray}
\frac{E}{\mu}&=&\frac{1-2v^2+\chi v^3}{\sqrt{1-3v^2+ 2\chi v^3}}\,,\\
\frac{L_z}{\mu}&=&\pm r_o v\frac{1- 2\chi v^3+\chi^2v^4}{\sqrt{1-3v^2+ 2\chi v^3}}\,,
\end{eqnarray}
where $v\equiv \sqrt{M/r_o}$ and the plus and minus sign correspond to prograde and retrograde orbits, respectively. 

Under the assumption that the evolution of the system under radiation reaction is adiabatic, i.e.~the radiation reaction timescale is much longer than the orbital period, we evolve the system using the balance equation:
\begin{equation}
\dot{E}\equiv \dot{r}_o\frac{dE}{dr}=-\dot{E}_{\rm GW}\,,
\end{equation}
where  $\dot{E}_{\rm GW}$ is the total GW flux. The evolution of the orbital phase $\phi$ can then be computed using 
\begin{equation}
\dot{\phi}=\Omega(t)\equiv \pm \frac{M^{1/2}}{r_o(t)^{3/2}+a M^{1/2}}\,.
\end{equation}
An example of the evolution of the orbital radius under radiation reaction when including the influence of tidal heating (i.e. when ${\cal R}=0$) is shown in Fig.~\ref{fig:radius} for 
$M=10^6M_\odot$, $\mu=30M_\odot$, and various spin values. The evolution starts at $r_o=10M$ up to the ISCO, so 
that in the highly-spinning case the evolution lasts longer. 

In the following we will be interested in computing the GW phase shift between an EMRI in a BH or ECO background. To 
do so we take into account the fact that the GW phase of the dominant mode is given by $\phi_{\rm GW}=2\phi$ and 
define the instantaneous dephasing as
\begin{equation}
\delta{\phi}(t)=\phi^{\rm BH}_{\rm GW}(t)-\phi^{\rm ECO}_{\rm GW}(t)\,,
\end{equation}
where $\phi^{\rm BH}_{\rm GW}(t)$ and $\phi^{\rm ECO}_{\rm GW}(t)$ denote the instantaneous GW phase in the BH and ECO case, respectively, and we have chosen the initial conditions such that $\phi^{\rm BH}_{\rm GW}(t=0)=\phi^{\rm ECO}_{\rm GW}(t=0)$ at the initial orbital radius of the evolution. 

In addition we also define the total dephasing accumulated up to a radius $r_f$ as
\begin{equation}
\Delta{\phi}=\phi^{\rm BH}_{\rm GW}(r_o=r_f)-\phi^{\rm ECO}_{\rm GW}(r_o=r_f)\,,
\end{equation}
where $\phi^{\rm BH,ECO}_{\rm GW}(r_o=r_f)$ is computed at the time where the orbital radius reaches $r_f$ and where again we set the orbital phase to be the same for both the BH and ECO case at the initial orbital radius.
\begin{figure*}[th]
\centering
\includegraphics[width=0.45\textwidth]{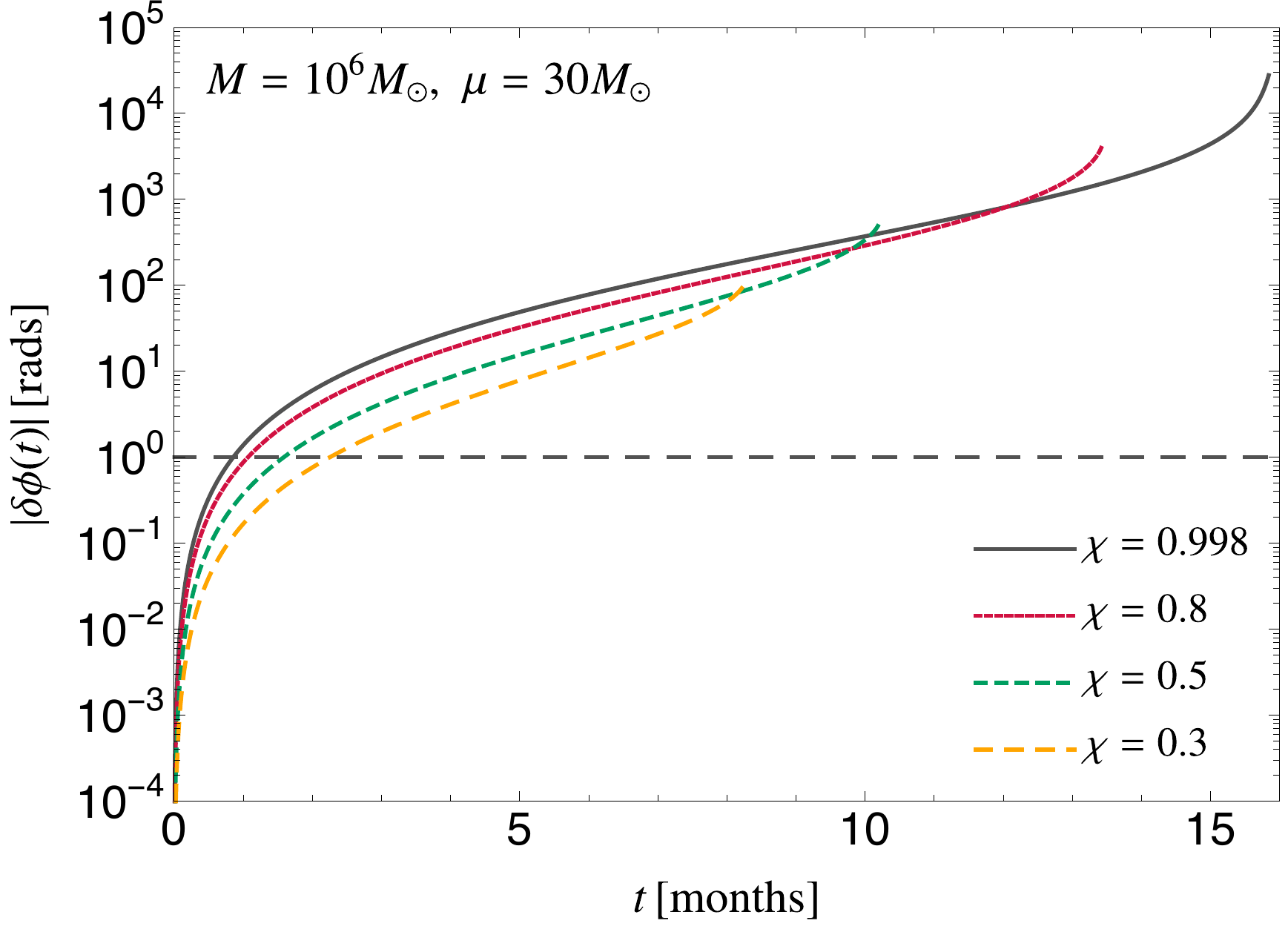}
\includegraphics[width=0.45\textwidth]{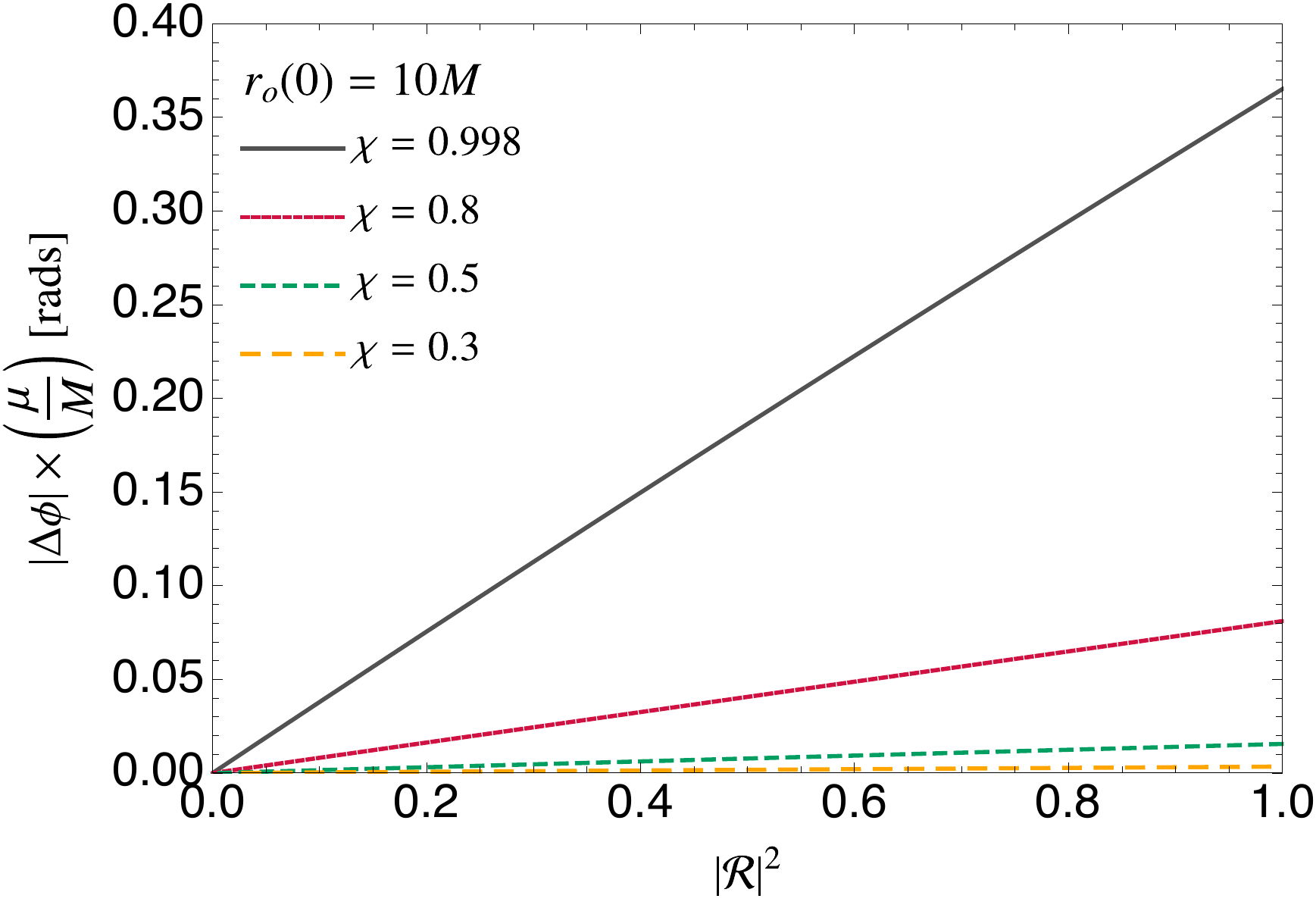}
\caption{Left panel: dephasing due to absence of tidal heating in a perfectly-reflective ECO ($|{\cal R}|^2=1$) relative to the BH case (${\cal R}=0$). 
We consider a prototypical EMRI system and the evolution starts at $r_o=10M$ up to the ISCO for various 
values of the spin of the central object.
Right panel: total dephasing accumulated from $r_o=10M$ up to the ISCO as a function of the 
reflectivity $|{\cal R}|^2$. 
Each line represents an interpolation of 20 equally spaced data points ranging from $|{\cal R}|^2=0$ up to $|{\cal 
R}|^2=1$. The dependence is linear for any value 
of the spin and of the other parameters.}
\label{fig:dephasing}
\end{figure*}

\subsection{Description of the code}

We integrate the perturbation equations and compute fluxes and waveforms using the Gremlin code available in the 
Black Hole Perturbation Toolkit~\cite{BHPToolkit}. 
This code uses an accurate continued-fraction representation of the solution to Teukolsky 
equation~\cite{Fujita:2009uz, Fujita:2004rb}. More specifically, the solutions $R^{\rm up, in}_{\ell m\omega}$ of the 
homogeneous Teukolsky equation are expanded as a series of hypergeometric functions; the coefficients of the series are 
determined by a three-term recurrence relation~\cite{Sasaki:2003xr}. 

We used Gremlin to solve Teukolsky equation for all $m$ modes up to $\ell=\ell_{max} = 20$ over a range of 
orbital radii. With the solutions for each $(\ell, m)$ mode at hand, the energy fluxes  can be computed by summing over 
all modes using Eq.~\eqref{EdotINF} and~\eqref{EdotH}, respectively. 
Thus, the fluxes are computed as a function of the orbital radius; data are evenly spaced in the range from $r=20 M$ to 
the ISCO, the latter depending on the value of the spin of the central object. At the ISCO, the fractional 
accuracy for the flux at infinity is $10^{-4}$ and for the flux at the horizon it is $10^{-8}$
\cite{Taracchini:2013wfa}. Finally, fluxes are used to evolve the 
orbital trajectory of the small body adiabatically starting at $r_o=10M$, together with the corresponding GW signal.

The adiabatic inspiral is driven by the energy loss from the orbit via GWs at infinity and tidal heating.
To compute the contribution of tidal heating we generated several sets of waveforms for different spin values.  One set of 
waveforms is constructed where the inspiral is driven by both tidal heating and GW emission to infinity. The other 
family of waveforms is constructed by considering different values for $|{\cal R}|^2$. We use these waveforms to 
calculate the mismatch as explained below.

\section{Results}

In the left panel of Fig.~\ref{fig:dephasing} we show the phase difference as a function of time for the case of a 
perfectly-reflective ECO ($|{\cal R}|^2=1$) and a BH with the same mass and spin. As a representative example, we use 
the same configurations as in Fig.~\ref{fig:radius}. The orbit is again evolved from $r_o=10M$ up to the ISCO, so that 
in the highly-spinning case the evolution lasts longer and the total dephasing is larger. 

The dashed horizontal line in the left panel of Fig.~\ref{fig:dephasing} marks the threshold $\delta \phi=1\,{\rm 
rad}$, which gives a very rough indication of the importance of tidal heating. As a rough but useful rule of thumb, if 
omission of tidal heating leads to dephasing $\delta\phi \approx 1\,{\rm rad}$ or greater as compared to a model
that includes tidal heating, then its omission is likely to substantially impact a matched-filter search, leading to a significant
loss of detected events~\cite{Lindblom:2008cm}.  We emphasize that this rule of thumb must be validated with a more careful analysis; for example, correlations
may allow for detection with incorrect models, albeit at the cost of systematic errors in fitted parameters.
As expected, both the total and the instantaneous dephasing grow with the spin~\cite{Hughes:2001jr}. In the example of 
Fig.~\ref{fig:dephasing}, for $\chi=0.3$, $\delta \phi\gtrsim 1\,{\rm rad}$ after slightly more than two months, 
whereas the same dephasing occurs after one month when $\chi\approx 0.9$. 
Overall, the total dephasing accumulated up to the ISCO is large, ranging from $10^2\,{\rm rad}$ to $10^4\,{\rm rad}$, 
depending on the spin.

In the right panel of Fig.~\ref{fig:dephasing} we show the dependence of the total dephasing with the reflectivity 
$|{\cal R}|^2$.  The dependence is linear, $\Delta\phi\propto|{\cal R}|^2$, to an excellent accuracy.
This is true also up to $|{\cal R}|^2\sim 1$, whereas the instantaneous dephasing $\delta \phi\propto |{\cal R}|^2$ 
only in the small-$|{\cal R}|^2$ or in the small-spin limit.
The $\Delta\phi\propto|{\cal R}|^2$ scaling 
allows us to compute the total dephasing for a single 
value of ${\cal R}$ and re-scale the final result for different values of the reflectivity a posteriori. For example, 
the dephasing for $|{\cal R}|^2=1/2$ would be approximately half of what shown in the left panel of 
Fig.~\ref{fig:dephasing}.

\begin{figure}[th]
\centering
\includegraphics[width=0.45\textwidth]{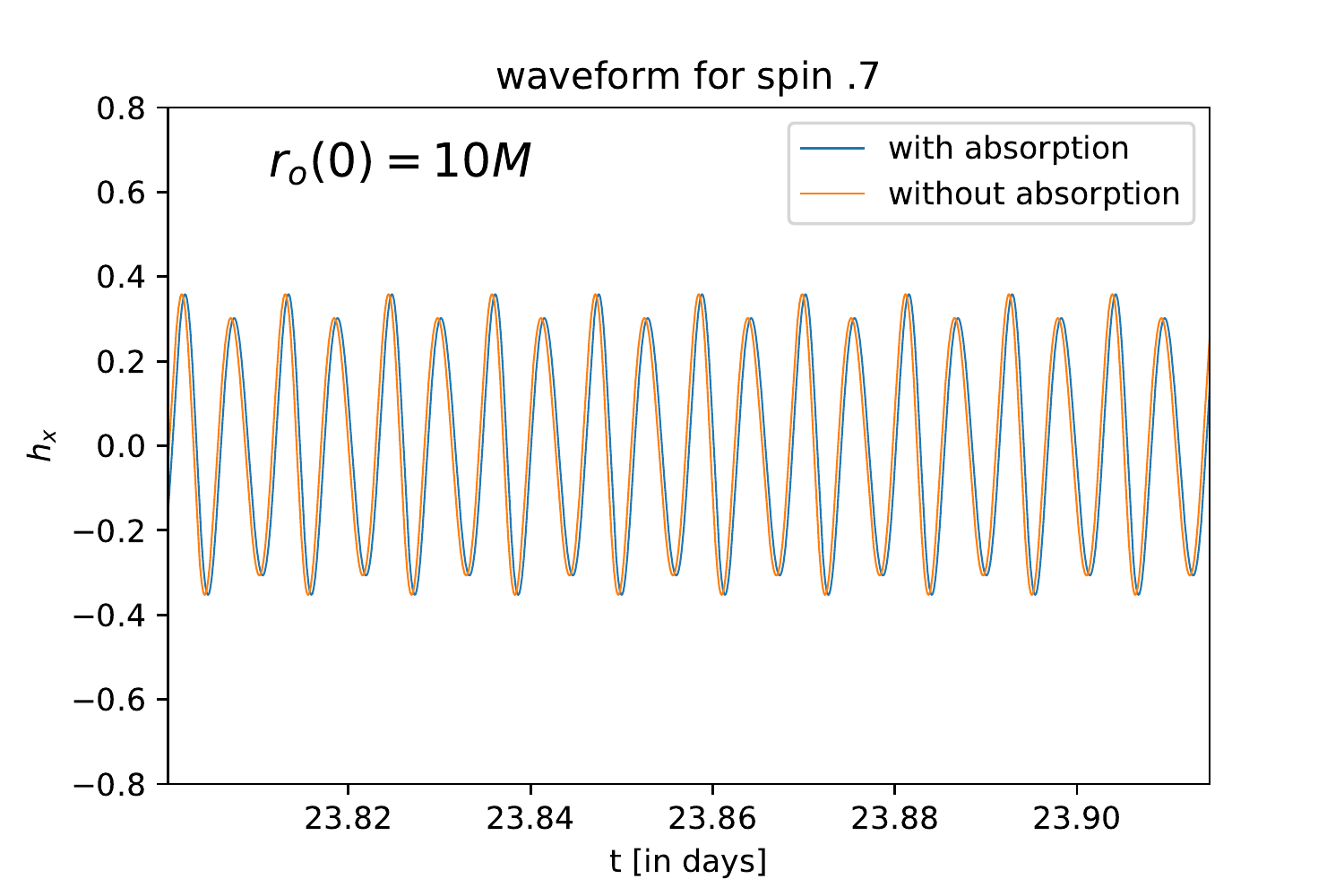}
\includegraphics[width=0.45\textwidth]{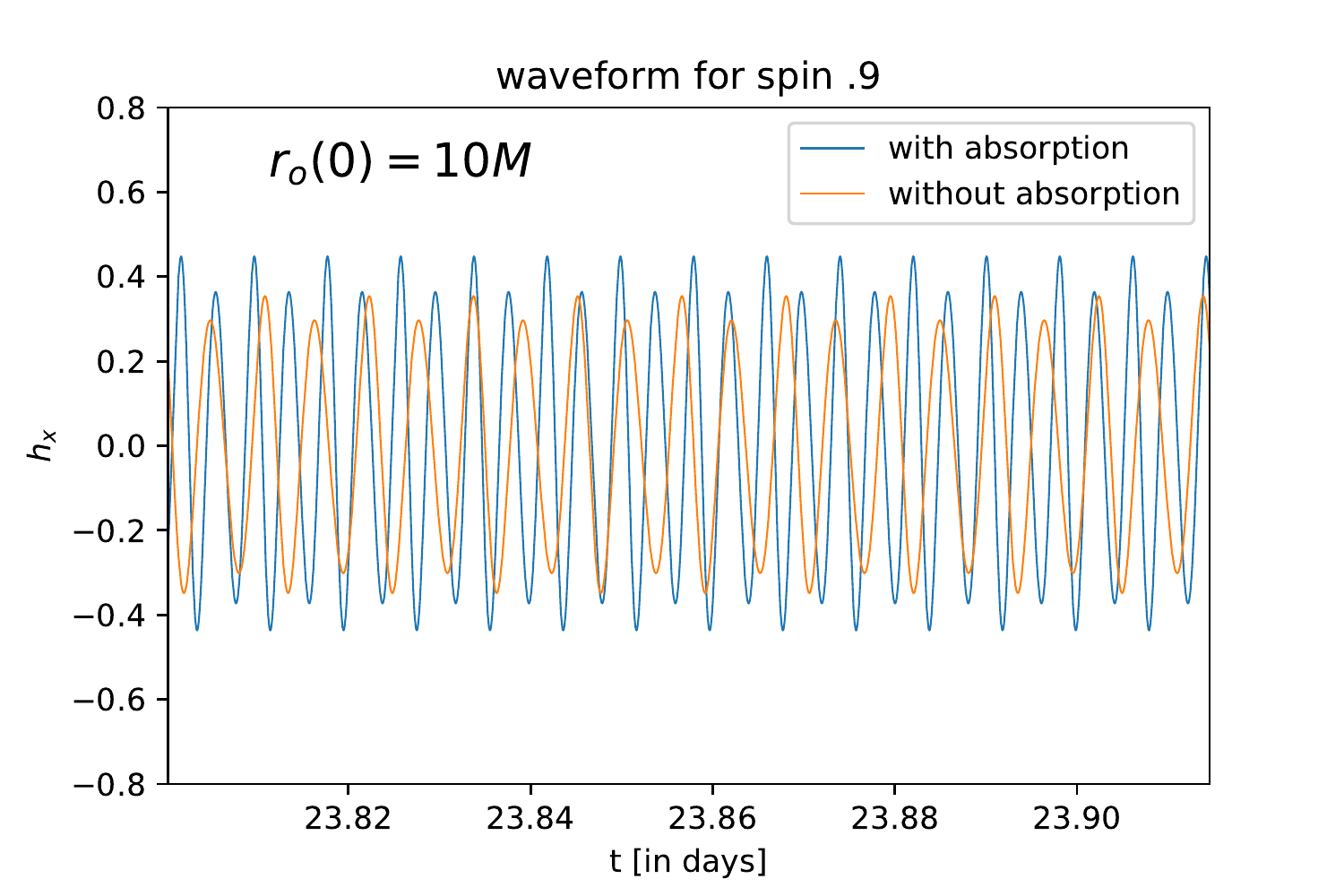}
\caption{Comparison between waveforms computed with $|{\cal R}|^2=0$ (with absorption) and $|{\cal R}|^2=1$ (without 
absorption). 
We show the waveforms as function of time for $M=10^6M_\odot$, $\mu=30M_\odot$ and an orbit with initial radius 
$r_o(0)=10M$ and an initial phase $\phi(0)=0$. We show a number of cycles roughly $23$~days after the beginning of the 
orbit, for $\chi=0.7$ (top) and $\chi=0.9$ (bottom). For larger spins the effect of tidal heating (i.e., of nonzero 
reflectivity at the surface of the central object) is more pronounced leading to larger dephasing between the 
waveforms. 
}
\label{fig:waveforms}
\end{figure}
\begin{figure*}[th]
\centering
\includegraphics[width=0.45\textwidth]{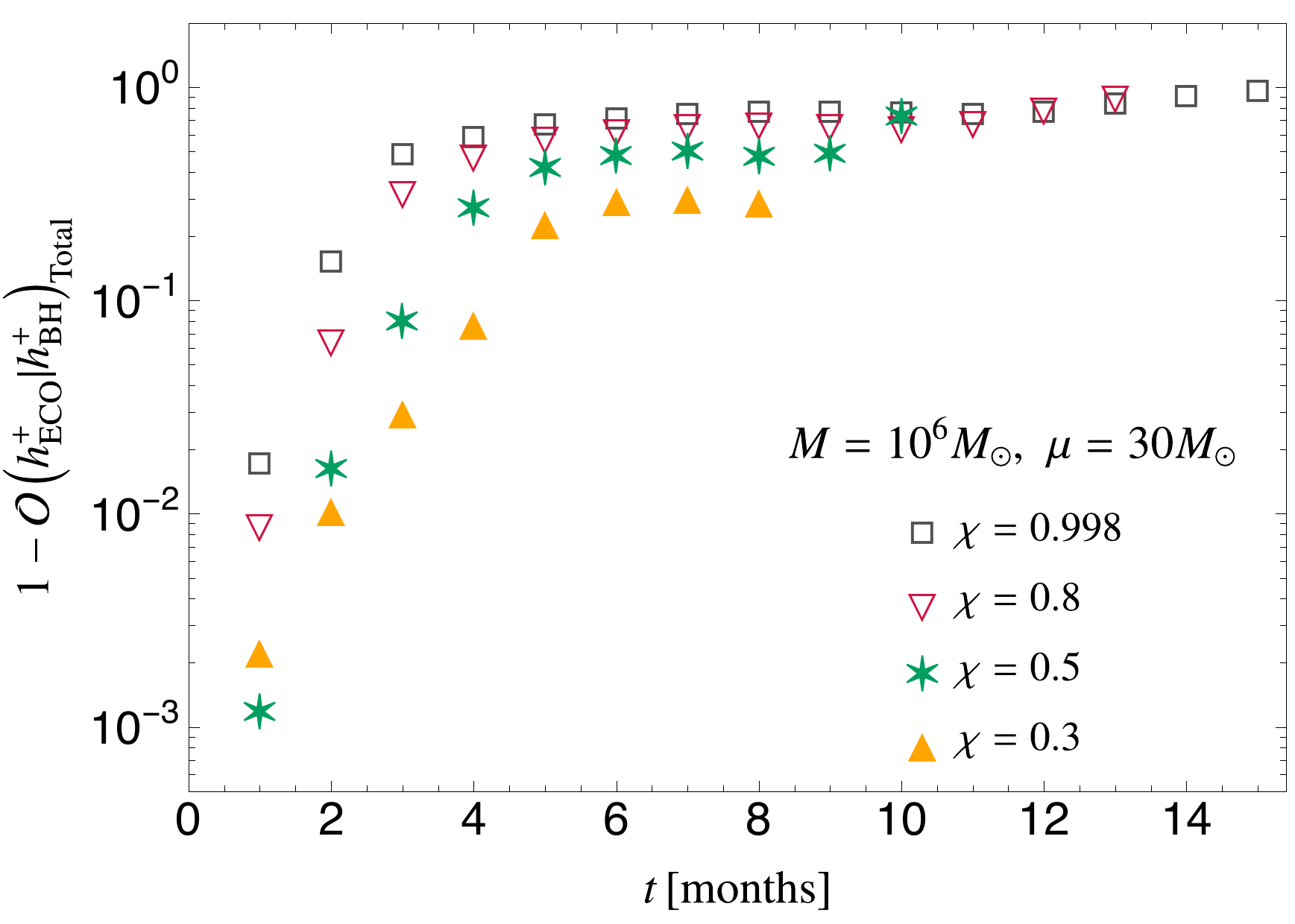}
\includegraphics[width=0.45\textwidth]{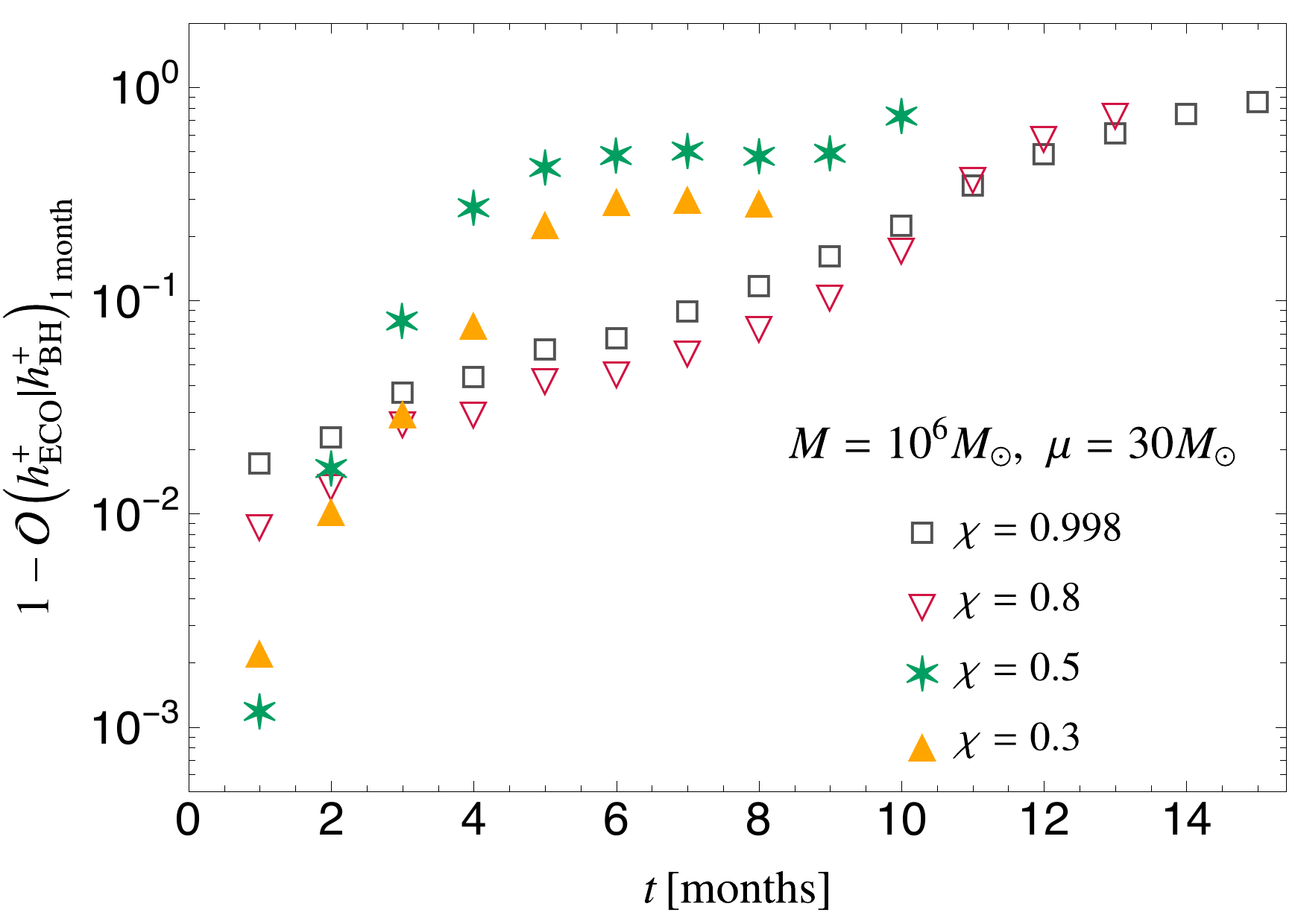}
\caption{Left panel: mismatch ${\cal M}=1-{\cal O}$ as a function of observation time between the plus polarization for 
a waveform computed with $|{\cal R}|^2=1$ (ECO) and another computed with $|{\cal R}|^2=0$ (BH), maximized over time 
and phase shift. We consider the same systems as in Fig.~\ref{fig:radius}.
Right panel: mismatch over chunks of one month at different stages of the evolution.
}
\label{fig:overlap}
\end{figure*}
\begin{figure}[th]
\centering
\includegraphics[width=0.48\textwidth]{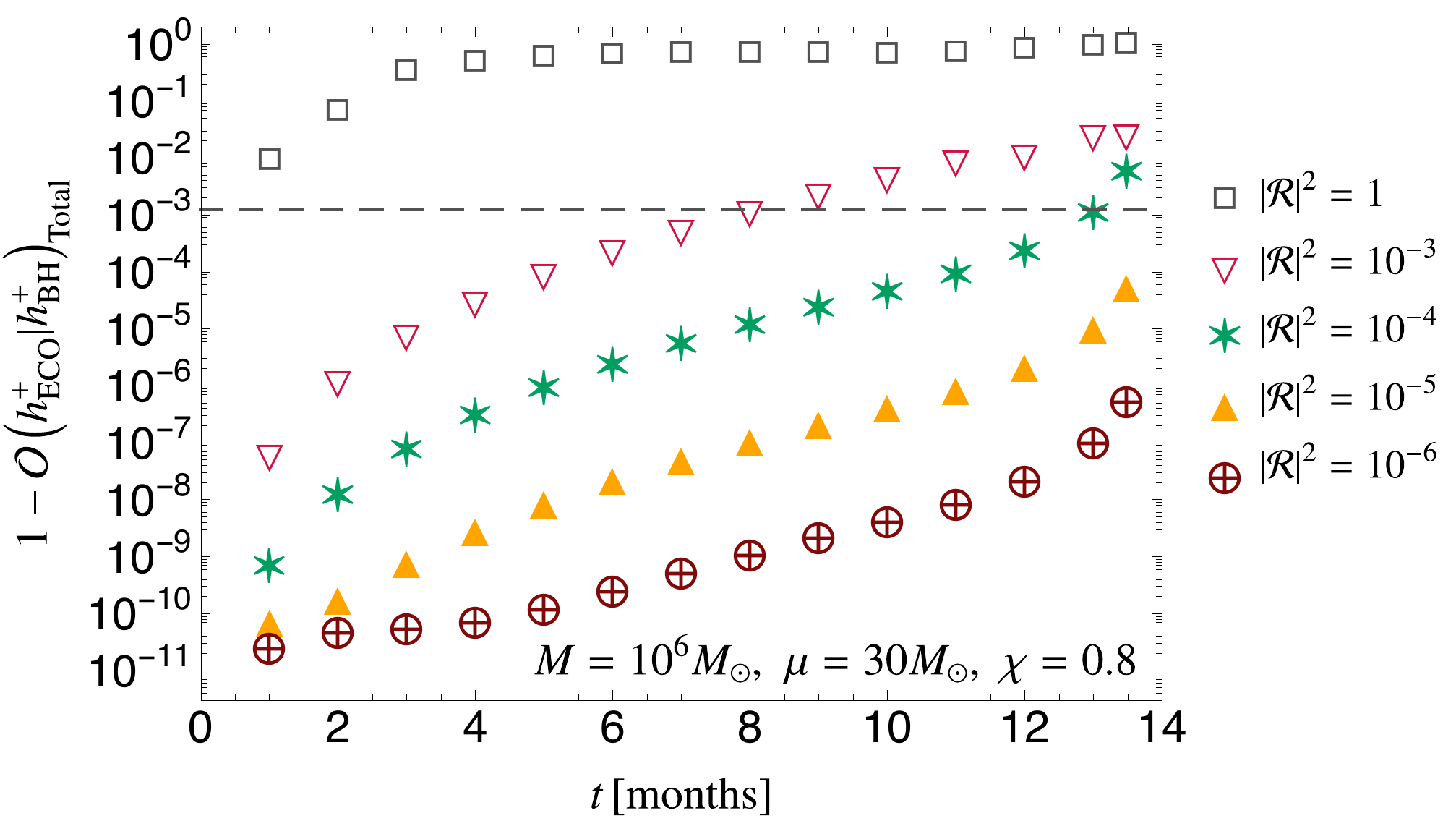}
\caption{Same as left panel of Fig.~\ref{fig:overlap} but also considering small $|{\cal R}|^2$ and for a central 
BH spin $\chi=0.8$. The dashed horizontal line marks the threshold $\mathcal{M}= 1/(2\rho^2)$, with $\rho=20$ being the 
fiducial SNR of the true signal.}
\label{fig:overlap_smallR}
\end{figure}

Note that absorption --~either total at the horizon or partial due to a partially reflecting ECO~-- also changes the 
mass and spin of the central object, in turn modifying the quasi-geodesic motion of the small orbiting body. This effect 
is always much smaller than the dissipative effect due to tidal heating. Even for a highly-spinning central object, this 
effect can account at most for a dephasing of $10^{-3}\,{\rm rad}$~\cite{Hughes_2019}, which is negligible compared to 
the effects shown in Fig.~\ref{fig:dephasing}.

In Fig.~\ref{fig:waveforms} we show two representative examples (for spin $\chi=0.7$ and $\chi=0.9$ in the top and 
bottom panels, respectively) of the GW waveform emitted during the EMRI to leading order in the mass ratio. 
Also in this case the effect of heating grows significantly with the spin and becomes even appreciable by the naked eye 
when $\chi\gtrsim0.9$.

Although the dephasing of the waveform is a useful measure to estimate the impact of tidal heating in the GW waveform, a 
better measure to assess whether the effect is sufficiently strong to be measurable in a GW detector with  noise power 
spectral density (PSD) $S_n(f)$, is to compute the overlap $\mathcal{O}$ between two waveforms $h_1(t)$ and $h_2(t)$:
\begin{equation}\label{overlap}
\mathcal{O}(h_1|h_2) = \frac{\left\langle h_1|h_2\right\rangle}{\sqrt{\left\langle h_1|h_1\right\rangle \left\langle h_2|h_2\right\rangle}}\,,
\end{equation}
where the noise-weighted inner product $\left\langle h_1|h_2\right\rangle$ is defined by
\begin{equation}
\left\langle h_1|h_2\right\rangle = 4\Re\,\int_{0}^{\infty} \frac{\tilde{h}_1 \tilde{h}^*_2}{S_n(f)} df\,.
\end{equation}
Here the tilded quantities stand for the Fourier transform and the star for complex conjugation. Since the waveforms 
are defined up to an arbitrary time and phase shift, it is also necessary to maximize the overlap~\eqref{overlap} over 
these quantities. In practice this can be done by computing~\cite{Allen:2005fk} 
\begin{equation}\label{overlap2}
\mathcal{O}(h_1|h_2) = \frac{4}{\sqrt{\left\langle h_1|h_1\right\rangle \left\langle h_2|h_2\right\rangle}}\max_{t_0} 
\left|\mathcal{F}^{-1}\left[\frac{\tilde{h}_1 \tilde{h}^*_2}{S_n(f)}\right](t_0)\right|\,,
\end{equation}
where $\mathcal{F}^{-1}[g(f)](t) =\int_{-\infty}^{+\infty} g(f) e^{-2\pi i f t}df$ represents the inverse Fourier 
transform. The overlap is defined such that $\mathcal{O}=1$ indicates a perfect agreement between the two waveforms. For the PSD we use the LISA curve of Ref.~\cite{Cornish:2018dyw} adding the contribution of the confusion noise from the unresolved Galactic binaries for a one year mission lifetime.

In Fig.~\ref{fig:overlap} we show the mismatch $\mathcal{M}\equiv 1-{\cal O}$ for the plus polarization of the waveforms 
with $|{\cal R}|^2=1$ and 
$|{\cal R}|^2=0$, for the systems considered in Fig.~\ref{fig:radius}. In the left plot of Fig.~\ref{fig:overlap} we 
show of the mismatch as a function of observation time for orbits starting at $r_o(0)=10M$.  For all the cases 
considered the mismatch  $\mathcal{M}<0.02$ until the first month of observation, however it quickly increases as the 
small object  approaches the ISCO, making the waveforms clearly distinguishable from one another.
In the right plot of Fig.~\ref{fig:overlap} we instead divide the waveforms in chunks of one month and compute the 
mismatch for that particular month of data. This allows us to assess how close are the waveforms at different stages of 
the evolution. 
As expected, the closer the object is from the ISCO the smaller the overlap. In particular for small spins the mismatch is 
$\mathcal{M}<0.1$ for most of the evolution. 
Finally, in Fig.~\ref{fig:overlap_smallR} we show how the mismatch depends on the reflectivity for small values of 
$|{\cal R}|^2$ for the system with $\chi=0.8$. As expected, the mismatch decreases with $|{\cal R}|^2$. We find that for 
$|{\cal R}|^2\lesssim 10^{-4}$ the mismatch behaves roughly as ${\cal M}\propto |{\cal R}|^4$. Indeed, in the small 
dephasing limit, ${\cal O}\propto \cos\delta\phi$~\cite{Lindblom:2008cm} and --~owing to the $\delta\phi\propto |{\cal 
R}|^2$ dependence~-- the mismatch should scale as ${\cal M}\propto |{\cal R}|^4$, in agreement with our 
results.\footnote{We note that, due to numerical errors in the waveforms, the ${\cal M}\propto |{\cal R}|^4$ 
scaling breaks down for very small mismatches, ${\cal M}\lesssim 10^{-10}$.}

\section{Discussion}

As a useful rule of thumb two waveforms are considered indistinguishable 
for parameter estimation purposes if $\mathcal{M}\lesssim 
1/(2\rho^2)$~\cite{Flanagan:1997kp,Lindblom:2008cm}, where $\rho$ is the SNR of the true signal. For an EMRI with an 
SNR $\rho\approx 20$ (resp., $\rho\approx 100$) one has $\mathcal{M}\lesssim 10^{-3}$ (resp., $\mathcal{M}\lesssim 
5\times 10^{-5}$). As a reference, in Fig.~\ref{fig:overlap_smallR} we mark the threshold $\mathcal{M}= 10^{-3}$ with a 
dashed horizontal line.
It is clear from Figs.~\ref{fig:overlap} and~\ref{fig:overlap_smallR} that this level of mismatch is quickly exceeded in an EMRI due to 
absence/presence of tidal heating for a perfecly-reflecting ECO~(i.e., when $|{\cal R}|=1$), even for small spins. This 
implies that the reflectivity $|{\cal R}|^2$ can be constrained down to very small values. For example considering a 
supermassive object with $\chi\gtrsim 0.8$ and a signal with $\rho=20$, from the results in 
Fig.~\ref{fig:overlap_smallR} we can estimate a very stringent bound on the reflectivity
\begin{equation}
 |{\cal R}|^2 \lesssim 5\times 10^{-5}\,. \label{bound}
\end{equation}
A more conservative bound would be obtained by requiring that the dephasing be smaller than $1\,{\rm rad}$. Owing to the $\delta\phi\propto|{\cal R}|^2$ 
dependence and considering also $\chi\gtrsim0.8$, we find the slightly weaker constraint $|{\cal R}|^2\lesssim 10^{-4}$. 
Thus, an EMRI detection is sensitive to an effective reflectivity of the central supermassive object as small as
$\sim {\cal O}(0.01\%)$ (as a reference, we remind that in the BH case the reflectivity is zero and 
that for a neutron star it is practically unity, even when accounting for dissipation~\cite{1971ApJ...165..165E}).

The above results confirm previous findings that advocated for the importance of tidal heating in standard EMRI 
waveforms~\cite{Hughes:2001jr,Taracchini:2013wfa}: heating needs to be modelled accurately in order 
not to introduce a large dephasing and systematic errors. In addition, we showed that the inclusion/absence of 
tidal heating can be used as a strong, model independent, discriminator for the presence of a horizon in the central 
supermassive object.

%
Compared to other types of observations, this is a very stringent bound.
For instance, in order to achieve a bound of the order of Eq.~\eqref{bound} at $2\sigma$ confidence level from a 
negative echo search in the ringdown of a comparable-mass binary merger, a SNR of 
${\cal O}(10^3)$ in the ringdown would be needed~\cite{Maggio:2019zyv}. Reaching $3\sigma$ confidence level for the same 
bound would require 
a SNR of ${\cal O}(10^4)$ in the ringdown, which is well beyond what is expected with LISA, even for the loudest 
mergers~\cite{Audley:2017drz} (although such loud signals might be possible with future 
extensions~\cite{Baibhav:2019rsa}).

Our analysis relies only on the modification of the fluxes at the leading order in the mass ratio, i.e., we included 
only the leading-order dissipative part of the self-force~\cite{Poisson:2003nc,Barack:2009ux}, neglecting conservative 
contributions and higher-order terms.
While conservative contributions and high-order terms are crucially important for parameter estimation, their
impact is not likely to be confused for that of tidal heating, since tidal heating effects are typically much stronger,
at least for realistic values of the 
spin and when ${\cal R}$ is not negligibly small. Thus, we expect that reliable constraints can be obtained by 
modelling (partial) absence of tidal heating in state-of-the-art waveform approximants to the leading 
order, along with --~and independently of~-- other self-force corrections.

We considered here the simplest trajectory, namely a circular equatorial orbit, but we expect that our 
results would remain qualitatively the same for more generic trajectories. Eccentric orbits can probe regions 
closer to the central object than in the circular case, so the effect of tidal heating may be expected to be even larger in 
that case. On the other hand, the relative effect of tidal heating on the orbit tends to be smaller for highly non-equatorial 
orbits~\cite{Hughes:2001jr}.

Another natural extension of our work concerns the role of resonances due to the excitation of 
low-frequency quasinormal modes which are ubiquitous for ECOs~\cite{Pani:2010em,Macedo:2013jja,Cardoso:2019rvt}. These 
resonances are very narrow and have been shown to produce a negligible effect in the nonspinning 
case~\cite{Cardoso:2019nis}. It would be interesting to include them in a spinning model and to investigate the possible 
existence of floating orbits, namely the possibility that for certain circular orbits the (negative) flux emitted to infinity can be compensated by a (positive, due to superradiance) flux at the horizon, in the case the latter is resonantly enhanced~\cite{Cardoso:2011xi}. If this condition occurs the orbits can be metastable and introduce a large dephasing. However, preliminary analysis shows that the effect of tidal heating 
discussed here should nonetheless be dominant.

Finally, we made the conservative assumption that the external geometry of the central object can be described by the 
Kerr metric. ECOs might display several multipolar deviations from Kerr, whose amplitude --~in the ultracompact 
regime~-- is bounded by regularity arguments~\cite{Raposo:2018xkf,Barcelo:2019aif}. These deviations affect also the 
conservative part of the EMRI dynamics to the leading order in the mass ratio and would introduce a further diagnostic 
for the presence of horizons, similarly to the case of ``bumpy'' 
BHs~\cite{Collins:2004ex,Vigeland:2009pr,Moore:2017lxy}.

\begin{acknowledgments}
This work makes use of the Black Hole Perturbation Toolkit~\cite{BHPToolkit}.
S.B. and S.D.  acknowledges financial support provided by Tata Trusts. SD 
would like to thank University Grants Commission (UGC), India, for financial support as senior research fellow.
R.B. acknowledges financial support from the European Union's Horizon 2020 research and innovation programme under the Marie Sk\l odowska-Curie grant agreement No. 792862.
P.P. acknowledges financial support provided under the European Union's H2020 ERC, Starting 
Grant agreement no.~DarkGRA--757480, and under the MIUR PRIN and FARE programmes.
S.A.H.~is supported by NSF Grant No. PHY-1707549 and NASA Grant No. 80NSSC18K1091.
The authors would like to acknowledge networking support by the COST Action CA16104 and 
support from the Amaldi Research Center funded by the MIUR program "Dipartimento di 
Eccellenza" (CUP: B81I18001170001).
S.D. would like to thank Sudhagar S., Rajorshi Chandra, Deepak N Bankar and Malathi Deenadayalan for useful discussions related to handling computational runs on the SARATHI cluster at IUCAA.
\end{acknowledgments}
%

\appendix

\section{Energy fluxes: ECOs vs BHs}\label{app:flux}

In this appendix we study the differences in the energy fluxes between a BH and an ECO.

\subsection{Energy flux at infinity}
Here we show that the energy flux at infinity due to a point particle in circular motion around a Kerr-like 
object is \emph{independent} (within numerical accuracy) of the boundary conditions at the surface of the object 
(modulo narrow resonances). As a 
by-product, the flux is the same for a BH and for an ECO. Our study extends that done in Ref.~\cite{Cardoso:2019nis}, in 
which  low-frequency perturbations of nonspinning objects were considered. Instead, we consider the case in which the 
spin of the object and the frequency of the perturbations are arbitrary.

Our starting point is Teukolsky's equation~\eqref{radial equation}. It is 
convenient to make a change of variables by introducing the Detweiler's 
function~\cite{1977RSPSA.352..381D,Maggio:2018ivz}
\begin{equation}
 \Psi = \Delta^{-1} \sqrt{r^2+a^2} \left[\alpha \
R_{lm\omega}+\beta \Delta^{-1} \frac{dR_{lm\omega}}{dr}\right]\,,\label{DetweilerX}
\end{equation}
where  $\alpha$ and $\beta$ are certain radial 
functions~\cite{1977RSPSA.352..381D,Maggio:2018ivz}.
By introducing the tortoise coordinate $x$ as in Eq.~\eqref{tortoise}, Teukolsky's master equation becomes
\begin{equation}
 \frac{d^2 \Psi}{dx^2}- V(r,\omega) \Psi= S \,. \label{master}
\end{equation}
where $ S$ is a source term and the final potential $V$ is defined, e.g., in Ref.~\cite{Maggio:2018ivz}.
The asymptotic behavior of the potential is $V\to -\omega^2$ as $x\to\infty$ and $V\to-k^2$ as $x\to-\infty$.
The functions $\alpha$ and $\beta$ can be chosen such that the resulting potential $V$ is purely
\emph{real}~\cite{1977RSPSA.352..381D,Maggio:2018ivz}. Although the choice of $\alpha$ and $\beta$ is not unique, 
${\Psi}$ evaluated at the asymptotic infinities ($x\to\pm\infty$) remains unchanged up to a phase. Therefore, 
the energy and angular momentum fluxes are not affected~\cite{MTB}.

As discussed in the main text, the solution to Eq.~\eqref{master} can be found in term of the Green's function as
\begin{equation}
 \Psi = \frac{\Psi_+}{W} \int_{-\infty}^x dx\, \Psi_-  S +  \frac{\Psi_-}{W} \int^{+\infty}_x dx\, \Psi_+  S\,,
\end{equation}
where $\Psi_\pm$ are two solutions of the homogeneous equation which satisfy the correct boundary conditions at 
infinity (for the plus sign) and near the object (for the minus sign), whereas 
$W=\frac{d \Psi_+}{dx} \Psi_- - \Psi_+ \frac{d \Psi_-}{dx}$ is their Wronskian.
Regardless of the nature of the central object, the boundary condition at infinity reads
\begin{equation}\label{eq:BCinf}
 \Psi\propto\Psi_+ \sim e^{i\omega x}\,.
\end{equation}

Given an object with reflectivity ${\cal R}$, the boundary condition near its surface ($x=x_0\to-\infty$) 
is~\cite{Mark:2017dnq}
\begin{equation}\label{eq:boundary}
 \Psi\propto\Psi_- \sim e^{-i k(x-x_0)} + {\cal R} e^{ik(x-x_0)}\,.
\end{equation}

As discussed in the main text the flux at infinity can be computed as  $\dot E_\infty\propto |\Psi(x\to\infty)|^2$, 
where
\begin{equation}
 \Psi(x\to\infty) =\frac{e^{i\omega x}}{W} \int_{-\infty}^{+\infty} dx \Psi_-  S \,.
\end{equation}

For a point particle in circular equatorial motion, the source term can be schematically written as
\begin{equation}
  S = A(\omega)\delta(x-x_o)+B(\omega)\delta'(x-x_o)\,, \label{sourcedelta}
\end{equation}
where $x_o=x(r_o)$ is the orbital radius in tortoise coordinates. Then, standard treatment~\cite{Martel:2005ir} leads 
to the following solution
\begin{equation}
 \Psi(x\to\infty) =\left. e^{i\omega x}\frac{\hat A(\omega)\Psi_-(r_o)+\hat 
B(\omega){\Psi_-'}(r_o)}{W}\right|_{\omega=m\Omega}\,, \label{psisol}
\end{equation}
where $\hat A$ and $\hat B$ are two functions of the frequency related to $A$ and $B$ in Eq.~\eqref{sourcedelta}.

Finally, one can solve numerically the homogeneous equation with boundary conditions given by 
Eqs.~\eqref{eq:BCinf} and~\eqref{eq:boundary} in order to evaluate $\Psi_-$ and the Wronskian, and using the explicit 
form of the source term for circular orbits. 
One can verify numerically that $\Psi(x\to\infty)$ appearing in Eq.~\eqref{psisol} does not depend on the value 
of ${\cal R}$ in Eq.~\eqref{eq:boundary}, at least within the numerical accuracy of our code. In particular, the 
energy flux at infinity is the same regardless the value of 
${\cal R}$, including the BH case (${\cal R}=0$).
This argument is valid far from possible resonances in the flux. These resonances correspond to the poles of the 
Wronskian $W$, which occur near the real axis in the complex plane. Since the fundamental quasinormal modes of an ECO 
have very small imaginary part~\cite{Pani:2009ss,Maggio:2017ivp,Maggio:2018ivz}, these resonances are extremely 
narrow~\cite{Pani:2010em,Macedo:2013jja} and their contribution to the dynamics is 
negligible~\cite{Cardoso:2019nis}. 

An alternative way to understand this result is the following. A point particle in circular motion emits monochromatic 
radiation. Part of the latter goes directly to infinity, contributing to $\dot E_\infty$ regardless of the boundary 
conditions near the central object. Another fraction of the radiation is either reflected by the potential barrier 
produced by the gravitational field of the object (both in the BH and in the ECO case) or partially reflected by the 
surface of the object (only in the ECO case). In both cases this radiation is reflected back at the same frequency and 
can therefore be efficiently re-absorbed by the orbting particle\footnote{Unless the orbital frequency matches that of 
a resonance.}, which therefore does not lose the corresponding 
energy. This occurs as long as $T_{\rm arr}\lesssim T_{\rm RR}$ [cf. Eqs.~\eqref{Tarr} and \eqref{TRR}], which is 
typically the case, as we showed in the main text.

\subsection{Energy flux at the ECO surface}

Here we show that the energy flux at the ECO surface can be expressed as a fraction of the energy flux at the horizon 
of a Kerr BH with the same mass and spin. In the ECO case, the boundary condition in Eq.~\eqref{eq:boundary} represents the 
sum of an ingoing wave and of an outgoing wave with relative amplitude ${\cal R}$. 

Assuming $|x_0|\gg M$, we can evaluate the flux as $x\to-\infty$. For the ingoing wave, this flux will be proportional 
to (the square of the absolute value of)
\begin{equation}
 \Psi_{\rm absorbed}(x\to-\infty) =\frac{1}{W} \int_{-\infty}^{+\infty} dx\, \Psi_+  S\,,   \label{PsiIN}
\end{equation}
where $\Psi_+$ is the solution which is regular at infinity, and it is the same for both the BH and the ECO cases. 
Notice that $\dot E_H\propto|\Psi_{\rm absorbed}(x\to-\infty)|^2$ is the energy flux at the horizon in the BH case. In 
the ECO case, there is an extra contribution due to the outgoing wave in Eq.~\eqref{eq:boundary}. The flux in this case 
will be proportional 
to (the square of the absolute value of)
\begin{equation}
 \Psi_{\rm reflected}(x\to-\infty) =\frac{{\cal R}}{W} \int_{-\infty}^{+\infty} dx\, \Psi_+  S\,. \label{PsiOUT}
\end{equation}
Notice that this contribution has the opposite sign in the flux, since it accounts for energy that crosses the 
object's surface in the opposite direction. Since 
$\Psi_+$ is independent of ${\cal R}$, the integral in Eqs.~\eqref{PsiIN} and \eqref{PsiOUT} is the same, so the ratio 
of the absorbed to reflected fluxes is
\begin{equation}
 \frac{\dot E_{\rm absorbed}}{\dot E_{\rm reflected}}=|{\cal R}|^2\,.
\end{equation}
Finally, since the two contribution have opposite sign, we obtain Eq.~\eqref{EdotECO} in the main text.

\bibliographystyle{apsrev4}
\bibliographystyle{utphys}
\bibliography{References.bib}
\end{document}